\documentstyle[preprint,aps]{revtex}

\begin{document}     
\baselineskip16pt
\title{
Formation, Evolution, and the Structure of Fronts Produced By 
Unsteady Injection of Highly Magnetized, Relativistic Flows}
\author{Amir Levinson 
\footnote{Present address: School of Physics
and Astronomy, 
Tel Aviv University, Tel Aviv 69978, Israel}\\
CRSR, Cornell University, Ithaca, NY 14853\\
and\\
Maurice H.P.M. van Putten\\
Math Department, MIT, Cambridge, MA 02139}
\maketitle
\begin{abstract}
We study the formation, evolution and structure of dissipative 
fronts produced by overtaking collisions of relativistic 
streams, with emphasis on strongly magnetized flows.  
The evolution of the system is followed using analytical approach
in the simple wave regime and numerical simulations in the 
non-simple wave regime, until a steady-state is reached.  The 
steady-state structure of the front is then examined by solving
the appropriate jump conditions.  The conversion of magnetic energy
into kinetic energy is parametrized in terms of the Alfv\'en 4-velocity
inside the front.  The implications for gamma-ray jets are briefly
discussed.\\ 
\mbox{$Keywords:$ }{galaxies:jets-MHD-relativity-shock waves}
\end{abstract}

\section{Introduction}
Highly magnetized, relativistic outflows are believed to play 
an important role in a variety of astrophysical systems, 
e.g., pulsar winds (Kennel \& Coroniti 1984; Gallant, et al. 1992), 
Galactic and extragalactic superluminal sources (Burns \& Lovelace 
1982; Begelman \& Li, 1994; Levinson \& Blandford 1996), gamma-ray 
bursts (Usov 1992; Waxman 1995; Thompson \& Duncan 1993; Smolsky \&
Usov 1996), and others.  Such flows are likely to be 
unsteady, giving rise to formation of shocks which can 
lead ultimately to the dissipation of some fraction of 
the bulk energy, either in the form of thermal energy or, through
shock acceleration (Blandford \& Eichler 1987) or magnetic 
reconnection, high energy particles with nonthermal 
distributions.  As demonstrated by Romanova \& Lovelace (1997) 
recently, the creation of dissipative fronts in Poynting flux 
jets may give rise to $\gamma$-ray flares as often seen in blazars.
			    
One important difference between unmagnetized and highly 
magnetized flows is that, in the
former case the sound speed is limited to $c/\sqrt3$, 
whereas in the latter case the Lorentz factor associated 
with the speed of propagation of a disturbance - the 
magnetosonic speed - can be much larger.  Consequently, 
the conditions required for the formation of strong 
shocks are expected to be more stringent in highly 
magnetized flows.  Even when the upstream flow is 
supermagnetosonic, so that strong shocks are created, 
only a relatively small fraction of the incoming flow energy goes 
to heat the downstream region; the remainder is used 
up to slightly compress the magnetic field 
downstream (Kennel \& Coroniti 1984).  The situation might be markedly 
different however in the presence of rapid magnetic field 
dissipation behind the shock.

Motivated by the above considerations, we study in
this paper the formation, evolution, 
and structure of fronts produced by unsteady injection of relativistic 
streams, with particular emphasis on flows in which the energy 
flux is dominated by electromagnetic fields (commonly referred 
to as high sigma flows). 

\section{Theory}
\subsection{Basic equations}
The stress-energy tensor of a perfectly conducting fluid can be written in terms of 
$u^{\alpha}$, the fluid 4-velocity, and the magnetic field 4-vector $b^{\alpha}=
\tilde{F}^{\alpha\beta}u_{\beta}$, where $\tilde{F}^{\alpha\beta}$ is the dual electromagnetic 
stress-tensor, as
\begin{equation}
T^{\alpha\beta}=rh^* u^{\alpha}u^{\beta}+p^*g^{\alpha\beta}-b^{\alpha} b^{\beta}.
\end{equation}
Here, $g_{\alpha\beta}$ denotes the metric tensor, $r$ is 
the fluid rest mass density, $h^*=h+b^2/r$ and $p^*=p+b^2/2$, 
where $p$ is the proper gas pressure,
$h$ the specific enthalpy and $b^2=b^{\alpha}b_{\alpha}$.
Under the assumption of ideal MHD, Ohm's law reduces to the condition 
$F^{\alpha\beta}u^{\beta}=0$.  The MHD equations can then be 
written in a divergence form as (van Putten, 1993)
\begin{eqnarray}
\nabla_{\alpha}T^{\alpha\beta}=0,\nonumber \\
\nabla_{\alpha}(r u^{\alpha})=0,\\
\nabla_{\alpha}(u^{[\alpha}b^{\beta]}+g^{\alpha\beta}c)=0,\nonumber 
\end{eqnarray}
where $c:=b_{\alpha}u^{\alpha}=0$ is a conserved constraint.  
Note that $\sqrt{4\pi} b_{\alpha}$ represents now the magnetic 
field measured in the fluid rest frame, the absolute 
value of which we denote here by $B$.  The above set of 
equations must be supplemented by an equation of 
state for the gas.  In regions of adiabatic flow where 
$u^{\alpha}\nabla_{\alpha}S=0$, $S$ being the entropy, the 
specific enthalpy is given by $h=1+\int {dp/r}$, with  $p=p(r)$. 

Using eq. (1), the energy flux carried by the flow can be written in the form
\begin{equation}
{\cal F}=rh\gamma^2{\bf v}(1+u_{A}^2)=rh\gamma_{A}^2\gamma^2{\bf v},
\end{equation}
where ${\bf v}$ is the fluid 3-velocity, $u^{\alpha}_A=
b^{\alpha}/\sqrt{rh}$ is the Alf\'ven 4-velocity with respect 
to the fluid rest frame, and $\gamma_A$ is the corresponding 
Lorentz factor.
				
\subsection{Simple waves} 
To simplify the analysis, we shall consider in the following 1D 
flow with velocity $u^{\alpha}=(\gamma,u,0,0)$, a polytropic equation
of state, $p(r)=Kr^{\Gamma}$, and magnetic field perpendicular 
to the fluid 3-velocity.  Using Maxwell equations together with 
the continuity equation it is readily shown that $B/r$ is conserved 
along stream lines (de Hoffmann \& Teller 1958).  Consequently, the 
proper magnetic pressure can be expressed as, $p_B=\kappa r^2$, 
with $\kappa =B^2/8\pi r^2$.  Thus, the problem reduces essentially 
to that of a 1D hydrodynamic flow with an effective equation 
of state $p^*(r)=Kr^{\Gamma}+\kappa r^2$.  Next, we define
\begin{equation}
c_s^2={1\over h^*}{\partial p^*\over \partial r}=
{\Gamma Kr^{\Gamma-1}+2\kappa r\over
1+{\Gamma\over\Gamma-1}Kr^{\Gamma-1}+2\kappa r}.
\end{equation}
Evidently, in the absence of magnetic fields, viz., $\kappa=0$, 
$c_s$ is simply the sound speed.  In the presence of very strong 
fields, such that the gas pressure is much smaller than the 
magnetic field pressure, $c_s$ reduces to the Alf\'ven speed 
$u_A/\gamma_A$.  If the entropy and $\kappa$ are taken to be 
constant throughout the flow, one can define the variables
$d\sigma=c_sdr/r$ and $\lambda=\sinh^{-1}(u)$.  
Equations (2) then yield (van Putten 1991)
\begin{equation}
{\partial \chi_{\pm}\over \partial t}\pm A_{\pm}(\sigma,\lambda)
{\partial\chi_{\pm}\over \partial x}=0,
\end{equation}
with $\chi_{\pm}=\sigma\pm \lambda$, and $A_{\pm}(\sigma,\lambda)
=(c_s\pm v)/(1\pm c_sv)$, where $v=u/\gamma$ is the fluid 3-velocity.  
Equation (5) implies that $\chi_{+}$ and $\chi_{-}$
are constant along the characteristics $dx/dt=A_{+}$ and 
$dx/dt=A_{-}$, respectively.  Note that $A_{\pm}$ is the relativistic 
sum and difference of $c_s$ and $v$.  In the case of simple waves 
one of the Riemann invariants is a constant.  Let us suppose that
$\chi_{-}=$ const.  Then, $A_{+}$ depends only on the remaining 
invariant, $\chi_{+}$, and it is clear from eq. (5) that 
$\chi_{+}$ is constant along straight lines with slopes $A_{+}$, 
implying that $A_{+}$ is simply the speed of propagation of 
a disturbance in the observer frame.  For weak disturbances for 
which $v\rightarrow 0$, we obtain $A_{+}\rightarrow c_s$, 
verifying that $c_s$ is indeed the speed of propagation of 
the disturbance in the fluid frame.      
													   
For illustration, we consider unsteady injection of 1D, highly 
magnetized fluid with the following initial and boundary conditions:
\begin{eqnarray}
\lambda(x,t)=\lambda_o=const;\ \ \ \ \ t\le 0,\nonumber \\
\lambda(x=0,t)=\lambda_b(t)=\lambda_o+t/\tau;\ \ \ \ t>0.
\end{eqnarray}
The general solution is then given by
\begin{equation}
\lambda(x,t)=\lambda_b(t');\ \ \ \ \ t'=t-{x\over A[\lambda_b(t')]},
\end{equation}
where the subscript (+) has been dropped to simplify the notation.  
In order to solve for $t'$, we must determine $A(\lambda)$.  Since 
we are merely interested in high sigma flows, we can, to a good 
approximation, neglect the gas pressure.  This corresponds to the limit 
$K=0$ in equation (4).  We then obtain $A(\lambda)=
\tanh(3\lambda/2-\lambda_o/2+\lambda_{Ao})$, where 
$\sinh(\lambda_{Ao})=u_{Ao}$ is the Alf\'ven 4-velocity of the initial flow
(at $t\le0$).  On substituting the later expression for 
$A(\lambda)$ into eq. (7), we obtain a transcendental equation 
for the retarded time $t'$, which can be solved numerically.  The 
solution thereby obtained describes a steepening wave propagating 
at a velocity $\tanh(\lambda_o+\lambda_{Ao})$, the relativistic 
sum of $v_{Ao}$ and $v$.  This wave will steepen into
a shock after time $t_s$ at some distance $x_s$ from the injection point.  
The point $(t_s,x_s)$ is a pole of the function $\lambda(x,t)$, and
can be found from the equation $d\lambda/dx=\infty$.  The result is
\begin{equation}
t_s={\tau\over3}\sinh(2\lambda_o+2\lambda_{Ao});\ \ \ 
x_s=ct_s\tanh(\lambda_o+\lambda_{Ao}).
\end{equation}
If we associate the flow acceleration time $\tau$ with the 
size of the injection region, $R$, (this might be, for 
example, the radius of a black hole or the inner boundary of 
an accretion disk from which a jet is ejected), then in the 
ultra-relativistic limit eq. (8) gives, 
$x_s= R \gamma_o^2\gamma_A^2/3$.  Note that in this limit 
$\gamma_o\gamma_A$ is just the Lorentz factor associated with 
the speed of propagation of the disturbance with respect to the 
injection frame, so that $c\tau (\gamma_o\gamma_A)^{2}$ is roughly 
the distance from the injection point at which two disturbances
separated by a time $\tau$ collide.  This constraint 
on the location at which the front is created may have important 
implications for the characteristics of the high-energy 
emission from relativistic jets (Levinson 1996).  In the case of
the Crab pulsar the above criterion (although the different geometry
is expected to somewhat alter this result) implies that for $R\sim 10^6$ 
cm any fluctuations of the pulsar wind are not likely to
steepen into traveling shocks upstream the standing shock (which is 
produced as a result of the interaction of the pulsar wind with the 
confining, slowly expanding SNR, and is located at a radius of 
about 0.1 pc), unless $\gamma_o\gamma_A<10^5$.

\subsection{The evolution and structure of adiabatic fronts}
The simple wave solution breaks down after a shock is formed 
(i.e., for $t>t_s$).  To study the evolution of the front in 
the non-simple wave regime, we have solved equations
(2) numerically.  A detailed account of the numerical method 
used is given in van Putten (1993).  To test the code, we have 
compared the numerical results with the analytic solution obtained 
above in the simple wave regime.  The agreement is found to be good.
													    
An example is shown in fig. 1.  As seen from this example, the 
collision of the initial and injected fluids leads ultimately 
to the creation of a front consisting of two shocks moving away 
from each other, and a contact discontinuity across which the 
total pressure is continuous.  The front propagates at
a speed (i.e., the speed of the contact discontinuity) intermediate 
between that of the initial and injected flows, and expands at 
a rate proportional to the relative velocity of the two shocks. 
Consequently, there is a net energy flow into the front, giving 
rise to a pressure buildup inside.  After the injection is 
terminated, the front reaches a steady-state whereby the net energy 
flow into the front is balanced by adiabatic cooling, owing to 
the expansion of the front.  It is then convenient to explore 
the steady-state structure of the front using the appropriate jump 
conditions.  Now, in the ideal MHD limit the specific magnetic 
field, $B/r$, must be continuous across the shock.  This 
will no longer be true in the presence of magnetic field dissipation,
which may result from e.g., magnetic reconnection in the front, (this would 
require magnetic field topology more complicated than that invoked above). 
A self consistent treatment of dissipative fronts is beyond the scope 
of this Letter.  In order to explore the effect of magnetic 
field dissipation on the parameters of the shocked fluid,
we suppose that some fraction of the magnetic field energy is 
being converted into particle energy inside the front by some 
unspecified mechanism.  We can then treat the magnetic 
field behind each shock as a free parameter which lies in the 
range between zero (although the magnetic field is not likely 
to drop well below the equipartition value) and the maximum 
value obtained in the ideal MHD limit.  

The problem can be most easily solved in the rest frame of 
the front.  In this frame the jump conditions decouple and 
reduce essentially to those of two single, independent shocks.  
Let subscript -(+) refers to fluid quantities leftward 
(rightward) of the contact discontinuity, and denote quantities 
inside the front by subscript $f$.  The jump conditions at the 
right (+) and left (-) shocks can then be written in the front frame as,
\begin{eqnarray}
& &r_{f\pm}=\gamma_{\pm}(1-v_{\pm}/v_{s\pm})r_{\pm},\nonumber \\
& &r_{\pm}h^*_{\pm}\gamma_{\pm}^2(1-v_{\pm}/v_{s\pm})-p^*_{\pm}
=r_{f\pm}h^*_{f\pm}-p^*_{f\pm},\\
& &r_{\pm}h^*_{\pm}\gamma_{\pm}^2v_{\pm}(v_{\pm}-v_{s\pm})
+p^*_{\pm}=p^*_{f\pm}.\nonumber
\end{eqnarray}
Here  $v_{s\pm}$ denote the shocks 3-velocity, $v_{\pm}$ are the 
3-velocities of the fluids outside the front (i.e., ahead of 
the shocks) with respect to the front frame, and $\gamma_{\pm}$ 
are the corresponding Lorentz factors.  Since we are merely interested in 
the strong shock limit, we shall restrict our analysis, in what 
follows, to the ultrarelativistic case, viz., $\gamma_{\pm}>>1$.  
In this limit we can approximate the enthalpy inside the front
as, $h_{f\pm}\simeq [\Gamma_{f\pm}/(\Gamma_{f\pm}-1)](p_{f\pm}/r_{f\pm})$.  
Assuming that the fluids upstream the shocks are cold (i.e., 
$p_{\pm}=0$), eqs. (9) yield for the shock velocities,
\begin{equation}
v_{s\pm}=\pm{(2+u_{Af\pm}^2)\Gamma_{f\pm}-2\over 2+\Gamma_{f\pm}u_{Af\pm}^2},
\end{equation}
and for the front pressure and temperature to the right (left) of 
the contact discontinuity,
\begin{eqnarray}
{p_{f\pm}\over r_{\pm}c^2\gamma_{\pm}^2}&=&{4\Gamma_{f\pm}
(\Gamma_{f\pm}-1)(1+u_{A\pm}^2)(1+u_{Af\pm}^2)
\over (2+\Gamma_{f\pm}u_{Af\pm}^2)(2\Gamma_{f\pm}-2
+\Gamma_{f\pm}u_{Af\pm}^2)},\nonumber \\
{kT_{f\pm}\over mc^2\gamma_{\pm}}&\equiv&{p_{f\pm}\over 
r_{f\pm}c^2\gamma_{\pm}}={2(\Gamma_{f\pm}-1)(1+u_{A\pm}^2)\over 
2+\Gamma_{f\pm}u_{Af\pm}^2}.
\end{eqnarray}
Note that the shocks velocities given by eq. (10) are 
independent of the magnetic field outside the front.
													   
In the ideal MHD limit, the requirement that the specific 
magnetic field should be continuous across the shock implies 
that the Alf\'ven 4-velocities ahead and behind the shocks are 
related by,
\begin{equation}
\Gamma_{f\pm}u_{Af\pm}^4
+[2(\Gamma_{f\pm}-1)+(\Gamma_{f\pm}-4)u_{A\pm}^2]u_{Af\pm}^2-2u_{A\pm}^2=0,
\end{equation}
where eqs. (10) and (11) have been used in deriving eq. (12).  In 
the weakly ($u_{A\pm}<<1$) and strongly ($u_{A\pm}>>1$) 
magnetized cases eq. (12) simplifies to $(u_{Af\pm}/u_{A\pm})^2
=1/(\Gamma_{f\pm}-1)$ and $(u_{Af\pm}/u_{A\pm})^2=(4
-\Gamma_{f\pm})/\Gamma_{f\pm}$, respectively.  On substituting 
the latter results into eqs. (10) and (11) we recover the shock 
velocity and downstream temperature of a perfectly 
conducting fluid, obtained earlier by e.g., Gallant et al. (1992);
$v_{s\pm}=\pm(\Gamma_{f\pm}-1)$, $kT_{f\pm}=
(\Gamma_{f\pm}-1)mc^2\gamma_{\pm}$, in the weakly 
magnetized case, and $(u_{s\pm}/u_{A\pm})^2=(4-\Gamma_{f\pm})/(8-4\Gamma_{f\pm})$, 
$kT_{f\pm}=[(2\Gamma_{f\pm}-2)/(4-\Gamma_{f\pm})]
mc^2\gamma_{\pm}$, in the highly magnetized case.  We see that in 
the strong shock limit considered here, a perfectly conducting 
front is heated to relativistic temperatures for any value of the magnetic 
field.  However, while in the case of weakly magnetized streams 
all of the bulk energy per particle incident into the front 
as measured in the front frame, $\gamma_{\pm}mc^2$, is 
converted into random energy (the average energy per particle 
in a relativistic gas is $kT/[\Gamma_{f\pm}-1]$), in the highly magnetized 
case only a small fraction of the total bulk energy carried by 
the flow outside the front, specifically the fraction carried by 
the particles $\sim \gamma_{A\pm}^{-2}$ (see eq. [3]), is thermalized; 
the remainder of the energy is used up to compress the magnetic 
field inside the front.  Consequently, collisions of highly magnetized 
flows in the ideal MHD limit lead to the formation of 
magnetically dominated fronts. 
													     
The situation could be markedly different if the front is 
highly dissipative.  Then, $u_{Af\pm}$
can deviate substantially from the maximum value given by 
eq. (12).  As seen from eq. (11), the energy per particle, 
$kT_{f\pm}/(\Gamma_{f\pm}-1)$, increases as $u_{Af\pm}$ 
decreases, ultimately approaching $mc^2\gamma_{\pm}\gamma_{A\pm}^2$ 
in the limit of rapid magnetic field dissipation, 
i.e., $u_{Af\pm}<1$.  (We have assumed that the front is 
adiabatic.  It is conceivable that it will become radiative 
at temperatures lower than that derived above if the synchrotron 
or inverse Compton cooling time is sufficiently short.  We 
shall not consider this possibility in this Letter.)  The 
shock velocities approach $v_{s\pm}=\pm(\Gamma_{f\pm}-1)$ in this limit.

Now, the velocity of the contact discontinuity and, hence, the velocities
upstream of each shock with respect to the front frame, $u_{\pm}$, 
are unknown a priori.  The requirement that the total pressure 
inside the front is continuous across the contact discontinuity 
surface yields, in the strong shock limit, the relation,
\begin{equation}
\gamma_{+}\simeq\eta\gamma_{-};\ \ \eta^2={r_{-}\Gamma_{f-}(1+u_{A-}^2)
(1+u_{Af-}^2)(2+\Gamma_{f+}u_{Af+}^2)
\over r_{+}\Gamma_{f+}(1+u_{A+}^2)(1+u_{Af+}^2)(2+\Gamma_{f-}u_{Af-}^2)}.
\end{equation}
Note that in the case of a symmetric front, which prevails when $r_{-}=r_{+}$, 
$p_{-}=p_{+}$, and $u_{A-}=u_{A+}$, eq. (13) gives $\eta=1$ as it should 
of course.
								
To complete the solution, we transform to the frame in which 
the fluid is injected (hereafter referred to as injection frame;  
this could be e.g., the rest frame of a star or a black hole
from which a jet is emerged), and in which the velocities of 
the initial and injected flows, denoted here by $u'_{+}$ and 
$u'_{-}$, are specified.  The proper mass density, pressure, and magnetic 
field are of course relativistic invariants and have the same 
values in the injection frame as in the front frame .  We 
consider the case $u'_{-}>>u'_{+}\ge0$, namely both the initial 
and injected fluids propagate in the same direction.  On 
transforming to the injection frame we obtain,
\begin{equation}
\gamma_{\pm}=\gamma_{\pm}'\gamma_c-u'_{\pm}u_c,
\end{equation}
with $u_c$ being the front 4-velocity with respect to the 
injection frame.  Equations (13) and (14) can be solved now to 
yield $\gamma_c$ and $\gamma_{-}$.  One finds
\begin{eqnarray}
\gamma_{c}&=&{\eta u'_{-}-u'_{+}\over \sqrt{(\eta u'_{-}-u'_{+})^2-(\eta \gamma'_{-}-\gamma'_{+})
^2}},\nonumber \\
\gamma_{-}&=&{\eta (\gamma'_{-}-u'_{-})-(\gamma'_{+}-u'_{+})\over 
\sqrt{(\eta u'_{-}-u'_{+})^2-(\eta \gamma'_{-}-\gamma'_{+})^2}}.
\end{eqnarray}
For initial fluid at rest ($u'_{+}=0$) the above solution 
simplifies to $\gamma_{-}\simeq (\gamma'_{-}/2\eta)^{1/2}$, 
and for relativistic initial fluid to $\gamma_{-}\simeq(\gamma'_{-}/
4\eta\gamma'_{+})^{1/2}$.  On substituting the latter results 
into eqs. (10) and (11) we obtain an expression for the front
quantities in terms of $u'_{\pm}$.  Evidently, for roughly 
symmetric fronts ($\eta$ order unity) the front temperature 
scales roughly as $(\gamma'_{-}/4\gamma'_{+})^{1/2}$.  In 
the asymmetric case, one side of the front will be heated 
to much higher temperatures than the other.  Heat conduction
across the contact discontinuity surface may then become 
important if the magnetic field is not completely transverse 
there.
													  
\section{Discussion}
A highly conducting front produced by overtaking collisions 
of strongly magnetized, relativistic flows should be magnetically 
dominated, and is not an efficient converter of bulk energy into 
random energy (i.e., thermal energy or nonthermal distributions 
of accelerated particles).  Much higher temperatures
can be attained in the presence of rapid magnetic field dissipation 
inside the front, that may arise from magnetic reconnection or 
plasma instabilities near the shocks.  A fraction of
the dissipated field energy is likely to be radiated away in the 
form of electromagnetic emission with relatively high efficiency, 
owing to synchrotron and inverse Compton cooling  of the energetic particles.
The latter process can be very effective in the presence of 
background radiation with roughly isotropic distribution, as 
is believed to be the case for Galactic and extragalactic jets 
(Dermer \& Schlickheiser 1993; Sikora et al. 1994; Ghisellini \& 
Madau 1996; Levinson \& Blandford 1996).
The radiative friction may also provide a mean for dissipating 
the magnetic field.  Such fronts may explain the transient emission 
observed in many of these sources (Romanova \& Lovelace 1997).  However, 
in order to be efficient radiator, the front must be created at 
not too large a distance from the putative black hole, depending 
on the intensity and spectrum of the external radiation, thereby imposing a 
constraint on the magnetosonic speed (see discussion following 
eq. 8). In the case of streams consisting of electron-positron 
plasma, the front may become loaded with $e^{\pm}$ pairs
as a result of effective pair production.  Frequent creation of 
such fronts may lead to a transition from magnetic to particle 
dominated jet (Levinson 1996), as seems to be suggested by 
observations (Blandford \& Levinson 1995).  Again, the location 
at which the front is created might be important; the fronts must 
form above the cooling radius, which depends upon the intensity 
and spectrum of the background radiation, in order to avoid 
catastrophic X-ray production and the disruption of the jet.
													    
We thank R.D. Blandford for useful comments.
A.L acknowledges support from NSF grants AST 91-19475 and AST 93-15375, 
NASA grant NAG 5-3097, and Alon Fellowship.

\newpage

\centerline{{\bf Figure Captions}}

\noindent{Figure 1.  Three epochs are shown in two examples of a 1D
simulation of the formation of a relativistically
hot, magnetized front in which the magnetic field
is everywhere transverse to the fluid flow.
Displayed are the temperature,
the specific magnetic field, and the fluid 3-velocity
in the reference frame co-moving
with the initial flow (ahead of the front). A burst of new outflow
from the left is modeled by the initial 4-velocity distribution
$u^x(x)=\sinh\{1+\tanh(10\frac{x_0-x}{L})\}$, where $x$ is the coordinate 
along the outflow, $x_o$ represents the injection point
and $L$ is the length of the displayed computational grid. 
The initial density ($r=1$), pressure ($P=0.025$) and Alf\'ven 4-velocity
[$u_A=0.5$ in (a) and $u_A=1$ in (b)]
are taken to be uniform.  As seen, the higher
Alf\'ven velocity in (b) results in a broader front.
Note the gradual steepening of the leading
shock front, which is more delayed in (b).
The left shock becomes steady at a very early stage in}
both examples shown.

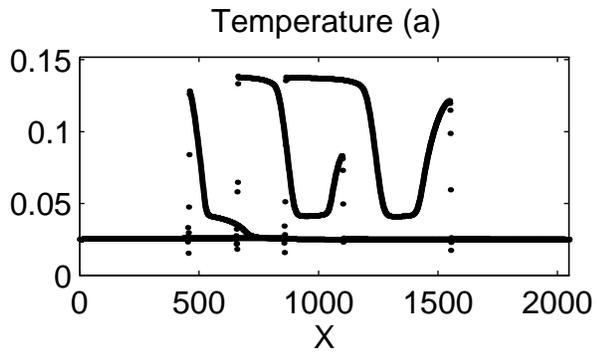
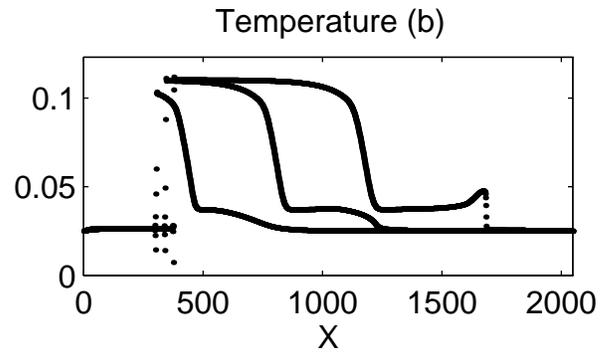
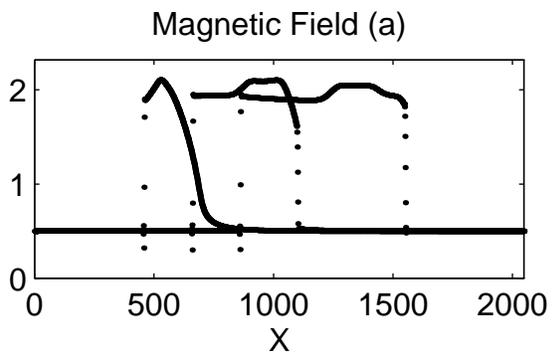
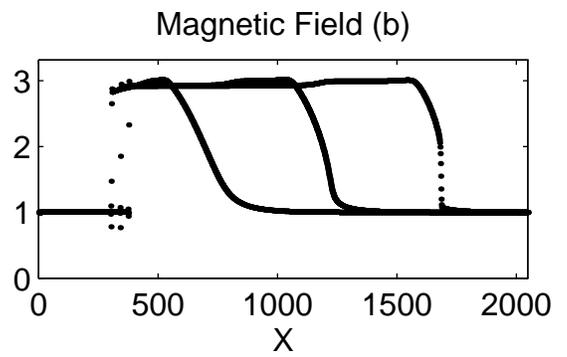
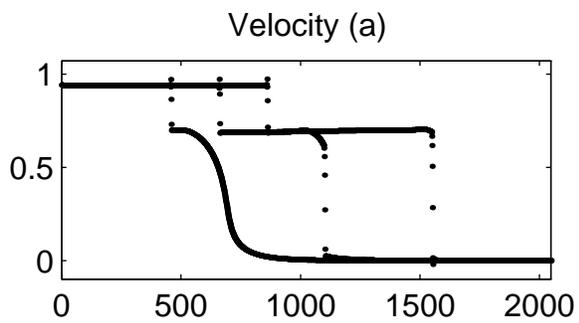
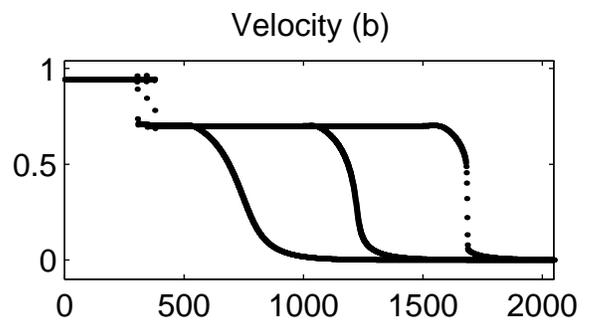
\end{document}